\documentclass{article}

\usepackage{PRIMEarxiv}

\usepackage{abstract}
\usepackage[utf8]{inputenc} 
\usepackage[T1]{fontenc}   
\usepackage{hyperref}       
\usepackage{url}            
\usepackage{booktabs}       
\usepackage{amsfonts}       
\usepackage{amsmath}        
\usepackage{amsthm}         
\usepackage{nicefrac}       
\usepackage{microtype}      
\usepackage{lipsum}
\usepackage{fancyhdr}       
\usepackage{graphicx}       
\usepackage{braket}       
\graphicspath{{media/}}     

\title{Introduction to QUDO, Tensor QUDO and HOBO formulations: Qudits, Equivalences, Knapsack Problem, Traveling Salesman Problem and Combinatorial Games
 }

\author{
  Alejandro Mata Ali \\
  Instituto Tecnológico de Castilla y León, Burgos, Spain\\
  \texttt{alejandro.mata.ali@gmail.com} \\
}

\begin{document}
\maketitle

\begin{abstract}
In this paper, we present a brief review and introduction to Quadratic Unconstrained D-ary Optimization (QUDO), Tensor Quadratic Unconstrained D-ary Optimization (T-QUDO) and Higher-Order Unconstrained Binary Optimization (HOBO) formulations for combinatorial optimization problems. We also show explicit encodings between these formulations and discuss their limitations. To help their understanding, we make some examples for the knapsack problem, traveling salesman problem and different combinatorial games. The games chosen to exemplify are: Hashiwokakero, N-Queens, Kakuro, Inshi no Heya, and Peg Solitaire. Although some of these games have already been formulated in a QUBO formulation, we are going to approach them with more general formulations, allowing their execution in new quantum or quantum-inspired optimization algorithms. This can be an easier way to introduce these more complicated formulations for harder problems.
\end{abstract}

\tableofcontents
\section{Introduction}
Combinatorial optimization has been one of the most studied and applied fields in recent years due to its application into industry. It involves determining a combination of elements \mbox{$\vec{x}=(x_0,x_1,\dots, x_{n-1})$} which minimizes a function $C(\vec{x})$, known as the \textit{cost function}. However, some of the most interesting problems are NP-Hard, without an efficient classical algorithm to solve them exactly. Interesting examples are the knapsack problem~\cite{knapsack} and the traveling salesman problem~\cite{TSP}, which are combinatorial optimization problems with use cases and easy to understand.

For this reason, one of the points of interest in recent quantum computing is combinatorial optimization. In this context, the most popular algorithm is \textit{Quantum Approximate Optimization Algorithm} (QAOA)~\cite{QAOA}. With this algorithm, an approximate solution to the minimization problem of a certain problem Hamiltonian can be obtained. A commonly used Hamiltonian is the \textit{Ising Hamiltonian} given its easy implementation, which consists of interactions of pairs of variables by means of $Z$ operators. For the treatment of classical combinatorial problems, the most commonly used formulation is the \textit{Quadratic Unconstrained Binary Optimization} (QUBO)~\cite{QUBO}, since it can be translated directly into an Ising Hamiltonian. The cost function of the QUBO problem is
\begin{equation}\label{eq: qubo function}
    C(\vec{x})=\sum_{i\leq j}^{n-1} Q_{ij}x_ix_j,
\end{equation}
being $x_i$ the $i$-th binary variable of the solution and $Q$ the matrix that defines the problem.

A straightforward generalization is the \textit{Quadratic Unconstrained D-ary Optimization} (QUDO), also called \textit{Unconstrained Quadratic Programming} (UQP). In this case, the variables are bounded nonnegative integers instead of binary, namely $x_i\in\{0,1,\dots,d_i-1\}$, or $x_i\in\{0,1,\dots,d-1\}$ in the constant-dimension case. Its equation is
\begin{equation}
    C(\vec{x})=\sum_{i\leq j}^{n-1} Q_{ij}x_ix_j + \sum_{i=0}^{n-1}D_ix_i,
\end{equation}
being $\vec{D}$ the vector that defines the linear part of the problem, because in this cause $x_i^2=x_i$ is not satisfied as in the binary case. When the natural values of a problem are $\{1,\dots,M\}$, we will distinguish them from the qudit labels by writing $y_i=x_i+1$. This formulation requires the use of qudits, a technology currently in study~\cite{Qudits,QAOA_qudit}. We can also generalize it even more with the \textit{Tensor Quadratic Unconstrained D-ary Optimization} (T-QUDO), with the pairwise categorical cost function
\begin{equation}
    C(\vec{x})=\sum_{i=0}^{n-1} U_i(x_i)+\sum_{i<j}^{n-1} V_{ij}(x_i,x_j),
\end{equation}
with $x_i\in\{0,1,\dots,d_i-1\}$. In the notation used later in the manuscript, the same information can be stored as
\begin{equation}
    C(\vec{x})=\sum_{i\leq j}^{n-1} Q_{i,j,x_i,x_j},
\end{equation}
being $Q$ the cost tensor which depends on two variable values and on which pair of variables we evaluate, with the diagonal blocks encoding the unary terms. This formulation has not been used in the state-of-the-art in quantum optimization with algorithms such as QAOA yet. Another possible generalization is to increase the order of the interactions. That is, instead of having a quadratic function, we have a multilinear pseudo-Boolean function of order at most $m$
\begin{equation}
    C(\vec{y})=\sum_{S\subseteq \{0,1,\dots,n-1\},|S|\leq m} q_S \prod_{i\in S} y_i,
\end{equation}
where $y_i\in\{0,1\}$. Because $y_i^2=y_i$, repeated indices can always be reduced to multilinear monomials. This is the \textit{Higher-Order Binary Optimization} (HOBO) formulation. This formulation has previously been studied~\cite{HOBO_rail} and is applicable to quantum devices~\cite{Opt_HOBO,QUBO_HOBO_TSP}.

All of these formulations are convenient to approach different combinatorial problems, but they may be more difficult to understand, formulate, and implement than the QUBO formulation. One particular way to introduce oneself into them is solving easier problems, maybe toy models or academic ones, and obtaining knowledge from them. The most interesting and easy case are the mathematical combinatorial games.

Mathematical games have always been of great interest due to their underlying mathematical properties~\cite{Hashi_NP}. This has led to their study to learn more about them and possible solving strategies. They are also useful for testing new ideas~\cite{Non_conflict_Queens,Survey_Queens}, technologies~\cite{Comparing_Nqueens,Quantum_Queens,Symbolic_Queens,GPU_Queens} and optimization algorithms~\cite{Bench_Hashi,Model_Peg}. Combinatorial games can be defined as those games in which there is only one player who knows all the information about the system, they are deterministic and the goal is to determine a combination $x$ that solves the problem and wins the game. This combination can be a configuration, in static games, or a combination of actions (sequence), in dynamic games. Previous work has presented QUBO formulations for Takuzu and N-Queens games~\cite{QUBOQueens}, indicating the susceptibility of this kind of problems to being formulated in this way. 

In this work, we present a brief introduction to QUDO, T-QUDO and HOBO formulations, exploring their limitations and uses. To understand them better, we create the formulations for the knapsack problem, the traveling salesman problem, the Hashiwokakero, the N-Queens, the Kakuro, the Inshi no Heya and the Peg Solitaire. This serves as examples of how to use these formulations. We optimize the number of variables needed and simplify the interactions.

\section{Quadratic Unconstrained D-ary Optimization (QUDO)}
As we have presented in the introduction, the Quadratic Unconstrained D-ary Optimization (QUDO) formulation consists in expressing a problem as an unconstrained combinatorial optimization problem with a cost function
\begin{equation}\label{eq: QUDO formulation}
    C(\vec{x})=\sum_{i\leq j}^{n-1} Q_{ij}x_ix_j + \sum_{i=0}^{n-1}D_ix_i,
\end{equation}
where $\vec{x}$ are bounded nonnegative integer variables, typically $x_i\in\{0,1,\dots,d_i-1\}$, and $Q$ and $\vec{D}$ are the matrix and vector which define the problem. If we want, we can transform this problem into a QUBO formulation, simply by the substitution of the QUDO variables $x_i$ with a set of binary variables $y_{ik}$
\begin{equation}
    x_i = \sum_{r=0}^{L_i-1} 2^r y_{ir},\qquad L_i=\lceil \log_2(d_i)\rceil,\qquad y_{ir}\in\{0,1\}.
\end{equation}
Substituting these expressions into Eq.~\eqref{eq: QUDO formulation}, we obtain a quadratic polynomial in the binary variables $y_{ir}$, so we recover a QUBO cost function of the form in Eq.~\ref{eq: qubo function}. If $d_i$ is not a power of two, the binary encoding introduces invalid codes with $x_i\geq d_i$; these codes must be removed with additional penalties or with a different encoding. Therefore, QUDO admits a standard encoding into QUBO, although this translation does not in general preserve resources.

In this case, the first thing to note is that we cannot take advantage of certain properties of the QUBO formulation, such as $x_i^2=x_i$. This forces us to introduce a linear term in the cost function to maintain its generality to the second order. The creation of the cost function does not have too much difficulty, as long as it is well adapted to this quadratic formalism, a fundamental condition for using this formalism as opposed to others. However, there are many constraints easily implementable in QUBO formulations that are not possible in QUDO, due to its higher dimensionality. We show several interesting cases:
\begin{enumerate}
    \item Counting constraint: in the QUBO formulation, if we want only a certain number $N$ of variables in $\mathcal{X}$ to take the value $1$, $\sum_{i\in\mathcal{X}} x_i=N$ we impose a restriction term $\left(N-\sum_{i\in\mathcal{X}} x_i\right)^2$. In the weighted case, $\sum_{i\in\mathcal{X}} a_ix_i=W$, we simply multiply each variable $x_i$ by its factor $a_i$, so the term is $\left(W-\sum_{i\in\mathcal{X}} a_i x_i\right)^2$. In the QUDO formulation, we can require that the sum of the values of the variables results in a particular value $N$, which is achieved with the same term. The same in the weighted case.
    
    If we want to determine that there are $N$ variables with non-zero values, i.e.\\ \mbox{$\sum_{i\in\mathcal{X}} H(x_i) = N$}, being $H$ the Heaviside step function, we need to have other type of terms, not compatible with the QUDO formulation. This is due to the fact that we do not know previously the value of the other variables. In the QUBO case, we know that if $N$ variables have non-zero values, they sum $N$, but in this case, they can have several different values.

    \item Inequality counting constraint: if we want the weighted sum of a set of variables to be less than a certain amount $Q$, in the QUBO formulation we use slack variables $\vec{s}$ in a way that $\sum_{i\in\mathcal{X}}a_i x_i\leq Q$ becomes $\sum_{i\in\mathcal{X}}a_i x_i+\sum_{j}2^js_j= Q$, requiring $L_s=\lceil\log_2(Q+1)\rceil$ slack variables (assuming $a_i\in \mathbb{Z}$). This is, the term is
    \begin{equation}
        \left(Q-\sum_{i\in\mathcal{X}} a_i x_{i}-\sum_{j=0}^{L_s-1}2^js_j\right)^2.
    \end{equation}
    Of course, you can avoid using slack variables with the unbalanced penalization~\cite{Unbalanced}, but we will keep them the rest of the paper for simplicity.
    
    In the QUDO case, we can impose the same restriction with a similar term. However, if the slack variables have $d$ possible values, we only need $L_s=\lceil\log_d(Q+1)\rceil$ slack variables and the constraint is $\sum_{i\in\mathcal{X}}a_i x_i+\sum_{j}d^js_j= Q$, with $s_j\in\{0,1,\dots,d-1\}$. The term is
    \begin{equation}
        \left(Q-\sum_{i\in\mathcal{X}} a_i x_{i}-\sum_{j=0}^{L_s-1}d^js_j\right)^2.
    \end{equation}
    The $Q+1$ is essential because the slack must represent every integer value from $0$ to $Q$ inclusive. For example, if $Q=8$, we need $\lceil\log_2(9)\rceil=4$ binary slack variables, not only $3$.
    However, if we want to do it for the number of non-zero variables, we have the same problem as before.
    
    \item Non-coincidence constraint: if two variables $x_i$ and $x_j$ cannot be non-zero at same time, in both QUBO and QUDO formulations, we impose a term $x_ix_j$. If we want to impose that they cannot be $x_i=a$ and $x_j=b$ at same time, i.e. $x_i= a \Longrightarrow x_j\neq b$ and $x_j= b \Longrightarrow x_i\neq a$, in QUBO we use
    \begin{equation}
        (1-(-1)^a(x_i-a))(1-(-1)^b(x_j-b)),
    \end{equation}
    where the $(-1)^a$ and $(-1)^b$ factors guarantee that their terms are the binary indicators of the events $x_i=a$ and $x_j=b$ for all possible values of $a$ and $b$. Explicitly, $I_0(x)=1-x$ and $I_1(x)=x$. In the QUDO case, we cannot impose in general this constraint, because it depends on the value of the sum when $x_j\neq b$, which can be different values if $x_j$ is not binary. This makes that the sign of the term depends on if the value of $x_j$ is higher or lower than $b$.

    \item Non-equality constraint: If we want them to satisfy $x_i\neq a \Longrightarrow x_j= b$ and $x_j\neq b \Longrightarrow x_i= a$, in the QUBO case we need a term
    \begin{equation}
        (-1)^{a+b}(x_i-a)(x_j-b).
    \end{equation}
    However, in the QUDO case, due to the same reason as before, the sign of the term is defined by the value of the $x_i$ and $x_j$ respect to $a$ and $b$, preventing this constraint to be implementable.
    \item Implication constraint: if we have a restriction $x_i=a \Longrightarrow x_j=b$, in the QUBO case we had a term
    \begin{equation}
        (1-(-1)^a(x_i-a))(-1)^b(x_j-b),
    \end{equation}
    so if $x_i=a$, the first term equals $1$, else equals $0$, and the second term is $0$ if $x_j=b$ and $1$ else. However, in the QUDO case there is the same problem with the sign of this term, making this constraint not implementable.
\end{enumerate}
We can see that in the QUDO formalism, the restrictions are hard to implement with respect to the QUBO case. However, this does not mean that this formulation is useless. We show a use case in the knapsack problem, reducing the amount of resources and variables, and in the Hashiwokakero, where the local bridge constraints can be completed with auxiliary variables into an exact QUDO formulation.

Before checking the examples, we can ask how the QUDO formulation can be implemented in the QAOA formulation. The answer is simple. We need to implement the qudit version of the QAOA~\cite{QAOA_qudit}, making use of the concepts presented in~\cite{Qudits} to create the cost hamiltonian gates. In this case, if we want a direct implementation, we do not need to implement the hamiltonian in an Ising-inspired way, like in the QUBO case. We can simply implement the exponential of the cost function operator, via qudit phase gates
\begin{equation}
P(\theta)=
    \begin{pmatrix}
        1&0&0&\cdots&0\\
        0&e^{i\theta}&0&\cdots&0\\
        0&0&e^{i2\theta}&\cdots&0\\
        \vdots&\vdots&\vdots&\ddots&\vdots\\
        0&0&0&\cdots&e^{i(d-1)\theta}
    \end{pmatrix},\quad P(\theta)_{kl}=\delta_{k,l}e^{i\theta k},
\end{equation}
for the linear terms, being a single qudit gate applied in $j$-th qudit with $\theta =\gamma D_j$, 
\begin{equation}
P2(\theta)=
    \begin{pmatrix}
        1&0&0&\cdots&0\\
        0&e^{i\theta}&0&\cdots&0\\
        0&0&e^{i2^2\theta}&\cdots&0\\
        \vdots&\vdots&\vdots&\ddots&\vdots\\
        0&0&0&\cdots&e^{i(d-1)^2\theta}
    \end{pmatrix},\quad P2(\theta)_{kl}=\delta_{k,l}e^{i\theta k^2},
\end{equation}
for the single variable quadratic terms, being a single qudit gate applied in $j$-th qudit with $\theta =\gamma Q_{jj}$, and
\begin{equation}
PP(\theta)_{l,m,n,o}=\delta_{l,n}\delta_{m,o}e^{i\theta lm},
\end{equation}
for the other quadratic terms, being a two-qudits gate applied in $j$-th and $k$-th qudits with $\theta =\gamma Q_{jk}$.

\subsection{Knapsack Problem}
The knapsack problem~\cite{knapsack} consists in choosing a set of elements to take into a knapsack, with the highest possible total value $V$ and without exceeding the total capacity $Q$. We have $N$ classes of objects, with a value $v_i$ and weight $w_i$ for the $i$-th class, which can be selected up to $c_i$ times. In the QUBO formulation, the problem is expressed with a set of variables $x_{ij}$, which indicate if we select the $j$-th object in the $i$-th class. This way, the problem is
\begin{equation}
    \begin{gathered}
        \text{maximize } V(x)=\sum_{i=0}^{N-1}\sum_{j=0}^{c_i-1} v_i x_{ij},\\
        \text{subject to } W(x)=\sum_{i=0}^{N-1}\sum_{j=0}^{c_i-1} w_i x_{ij} \leq Q.
    \end{gathered}
\end{equation}
In a QUBO cost function we can use
\begin{equation}
    C(x,s)=-\sum_{i=0}^{N-1}\sum_{j=0}^{c_i-1} v_i x_{ij}+\lambda \left(Q-\sum_{i=0}^{N-1}\sum_{j=0}^{c_i-1} w_i x_{ij}-\sum_{k=0}^{L_s-1}2^ks_k\right)^2,
\end{equation}
being $L_s=\lceil\log_2(Q+1)\rceil$ and $s_k\in\{0,1\}$ the slack variables. This requires $\sum_{i=0}^{N-1}c_i +L_s$ binary variables. A conservative sufficient condition is to choose $\lambda$ larger than the maximum objective gain obtainable from violating the capacity constraint. For nonnegative values, one convenient bound is
\begin{equation}
    \lambda>\sum_{i=0}^{N-1} c_i v_i,
\end{equation}
assuming that every infeasible integer violation contributes at least one unit to the squared penalty. This bound is sufficient, but not tight in general.

We can optimize the formulation with a condensed binary encoding of the multiplicities. Define
\begin{equation}
    x_i=\sum_{j=0}^{L_i-1}2^j x_{ij},\qquad L_i=\lceil\log_2(c_i+1)\rceil,
\end{equation}
so that $x_i$ represents the number of selected items of class $i$. If $c_i+1=2^{L_i}$, this encoding has no unused codewords; otherwise, the configurations with $x_i>c_i$ must be penalized or excluded. Then the problem becomes
\begin{equation}
    \begin{gathered}
        \text{maximize } V(x)=\sum_{i=0}^{N-1}\sum_{j=0}^{L_i-1} v_i 2^jx_{ij},\\
        \text{subject to } W(x)=\sum_{i=0}^{N-1}\sum_{j=0}^{L_i-1} w_i 2^jx_{ij} \leq Q,
    \end{gathered}
\end{equation}
being $x_{ij}=1$ if we activate the $2^j$ contribution of the $i$-th class. In this case, the QUBO formulation is
\begin{equation}
    C(x,s)=-\sum_{i=0}^{N-1}\sum_{j=0}^{L_i-1} v_i 2^jx_{ij}+\lambda \left(Q-\sum_{i=0}^{N-1}\sum_{j=0}^{L_i-1} w_i 2^jx_{ij}-\sum_{k=0}^{L_s-1}2^ks_k\right)^2,
\end{equation}
requiring $\sum_{i=0}^{N-1}L_i+L_s$ binary variables.

In the QUDO case, if we have qudits with the correct dimension $c_i+1$ for the main variables, and qudits of dimension $d$ for the slack variables, we can express the problem as
\begin{equation}
    \begin{gathered}
        \text{maximize } V(x)=\sum_{i=0}^{N-1} v_i x_{i},\\
        \text{subject to } W(x)=\sum_{i=0}^{N-1} w_i x_{i} \leq Q,
    \end{gathered}
\end{equation}
being $x_i\in\{0,1,\dots,c_i\}$ the number of times we choose an element of the $i$-th class.

The cost function in the QUDO formulation is
\begin{equation}
    C(x,s)=-\sum_{i=0}^{N-1} v_i x_{i}+\lambda \left(Q-\sum_{i=0}^{N-1} w_i x_{i}-\sum_{k=0}^{L_s^{(d)}-1}d^ks_k\right)^2,
\end{equation}
where $L_s^{(d)}=\lceil\log_d(Q+1)\rceil$ and $s_k\in\{0,1,\dots,d-1\}$. This requires $N+L_s^{(d)}$ variables. This is the opposite process described before, but now we have to get the QUDO from the QUBO. We can relax the dimensionality condition of the qudits mixing formulations. If we have qudits of dimension $d$ exclusively, we can encode each multiplicity as
\begin{equation}
    x_i=\sum_{j=0}^{L_i^{(d)}-1} d^j x_{ij},\qquad L_i^{(d)}=\lceil\log_d(c_i+1)\rceil,
\end{equation}
with $x_{ij}\in\{0,1,\dots,d-1\}$. If $c_i+1=d^{L_i^{(d)}}$, there are no unused codes; otherwise, values with $x_i>c_i$ must again be excluded. The problem can then be expressed with the cost function
\begin{equation}
    C(x,s)=-\sum_{i=0}^{N-1}\sum_{j=0}^{L_i^{(d)}-1} v_i d^jx_{ij}+\lambda \left(Q-\sum_{i=0}^{N-1}\sum_{j=0}^{L_i^{(d)}-1} w_i d^jx_{ij}-\sum_{k=0}^{L_s^{(d)}-1}d^ks_k\right)^2,
\end{equation}
being $x_{ij}$ the number of times we choose a group of $d^j$ elements of the $i$-th class. This requires $\sum_{i=0}^{N-1}L_i^{(d)}+L_s^{(d)}$ qudits. This formulation needs fewer variables than the QUBO one to give the same cost function.

\subsection{Hashiwokakero}
\begin{figure}[h]
    \centering
    \includegraphics[width=\linewidth]{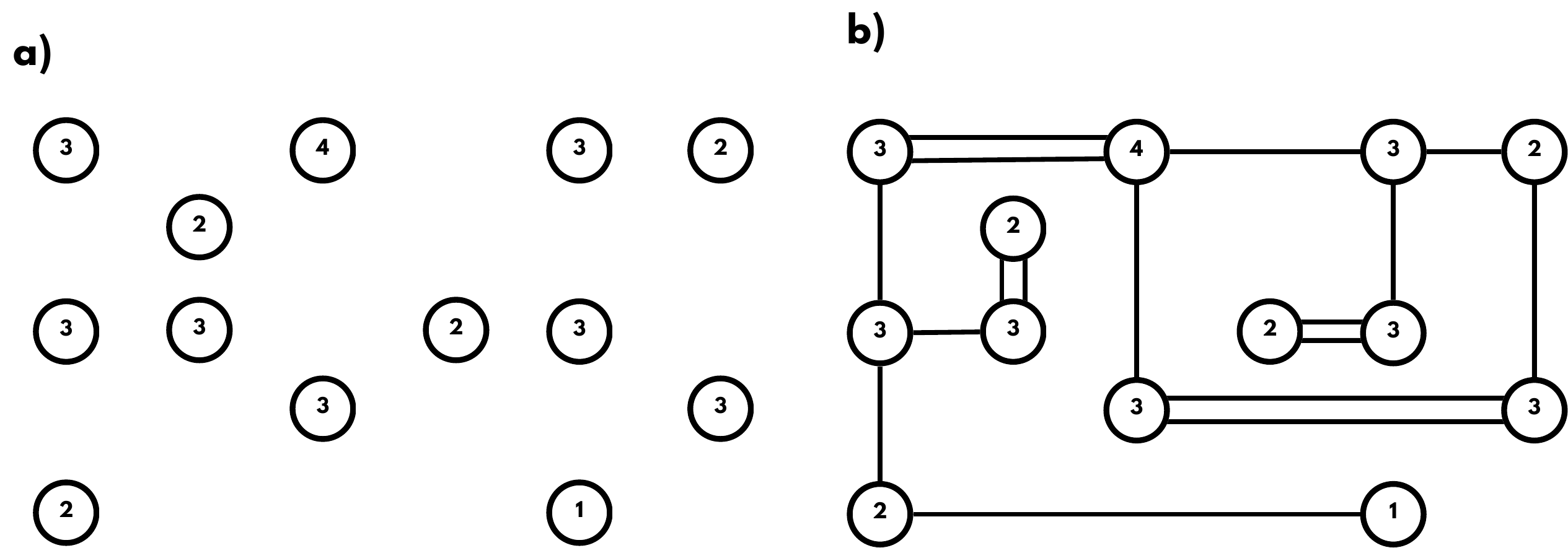}
    \caption{Hashiwokakero with a $5\times 6$ grid. a) Problem instance, b) Solution.}
    \label{fig:hashiwokakero}
\end{figure}

In this game, we have a grid of $N\times N$. In the set of vertexes $\mathcal{V}$ there are some nodes with numbers that indicate the number of edges that must connect them to their vertical and horizontal neighbor nodes. The node in position $(i,j)$ has a number $D_{ij}$. An example of the problem is shown in Fig.~\ref{fig:hashiwokakero} a. Also, each node can be connected to each other with zero, one, or two edges, and edges cannot cross each other. Finally, it can never be regions of vertexes isolated from others. That is, all islands must belong to a single connected component. A solution is shown in Fig.~\ref{fig:hashiwokakero} b. We need to determine the connections between nodes to ensure that every node has its corresponding number of connections.

As the variables to describe a solution naturally have three possible values, it is susceptible to be treated with a QUDO formulation. To properly formulate the problem in a simple way, we redefine it in graph terms. Let $V=\mathcal{V}$ be the set of islands and $n=|V|$. For each island $i\in V$, let $D_i$ be the objective number of connections of the $i$-th island. Let $E$ be the set of candidate bridge positions, where an edge $e=\{i,j\}\in E$ exists when the islands $i$ and $j$ can be connected by a horizontal or vertical bridge without passing through another island. Equivalently, $\mathcal{C}_i=\{j:\{i,j\}\in E\}$. In the standard puzzle, the variable $x_e$, or equivalently $x_{ij}$, indicates the number of connections between the $i$-th island and the $j$-th island, so $x_e\in \{0,1,2\}$. The connections between $(i,j)$ and $(j,i)$ are the same, so $x_{ij}=x_{ji}$. In addition, each island cannot connect to itself, so all the $x_{ii}=0$. Therefore, we only need one variable for every possible connection, avoiding impossible connections. As every island can have at most four candidate connections and half of them are repeated, we only need at most $2n$ local bridge variables with dimension three. The complete formulation below adds auxiliary flow and slack variables to enforce connectivity exactly.

The local part of the cost function consists of the degree term
\begin{equation}
    C_{\mathrm{deg}}=\lambda_d\sum_{v\in V}\left(D_v-\sum_{e\in\delta(v)}x_e\right)^2,
\end{equation}
and the non-crossing term
\begin{equation}
    C_{\mathrm{cross}}=\lambda_\times\sum_{\{e,e'\}\in \mathcal{X}} x_e x_{e'},
\end{equation}
being $\delta(v)$ the set of candidate edges incident to $v$ and $\mathcal{X}$ the set of unordered pairs of candidate edges that geometrically cross. The first term imposes the degree condition and the second forbids two crossing candidate bridges from being active at the same time.

These local terms are not enough to impose the final Hashiwokakero rule, namely that all islands must belong to a single connected component. To enforce that condition exactly, we introduce an auxiliary single-commodity flow, as commonly done in exact Hashiwokakero models~\cite{Bench_Hashi}. Let $r\in V$ be an arbitrary root island and define $B=n-1$. For each undirected candidate edge $e=\{i,j\}\in E$, we introduce two directed integer flow variables $f_{ij},f_{ji}\in \{0,1,\dots,B\}$. These auxiliary variables do not represent bridges; they only certify connectivity. We also define
\begin{equation}
    b_v=
    \begin{cases}
        n-1, & v=r,\\
        -1, & v\neq r.
    \end{cases}
\end{equation}
Then the flow-balance penalty is
\begin{equation}
    \begin{aligned}
        C_{\mathrm{bal}}=\lambda_b\sum_{v\in V}
        \left(
            \sum_{u:\{v,u\}\in E} f_{vu}
            -\sum_{u:\{v,u\}\in E} f_{uv}
            -b_v
        \right)^2.
    \end{aligned}
\end{equation}
This means that the root sends $n-1$ units of flow and every other island consumes one unit.

The flow must only pass through edges with at least one bridge. Therefore, for each $e=\{i,j\}\in E$ we require $f_{ij}+f_{ji}\leq Bx_e$. We convert this inequality into a QUDO equality with a slack variable $s_e\in \{0,1,\dots,2B\}$ and the penalty
\begin{equation}
    C_{\mathrm{cap}}=\lambda_c\sum_{e=\{i,j\}\in E}\left(Bx_e-f_{ij}-f_{ji}-s_e\right)^2.
\end{equation}
In the standard puzzle, the range $\{0,1,\dots,2B\}$ is sufficient because $x_e$ can take the values $0$, $1$ and $2$, so the capacity $Bx_e$ can be $0$, $B$ and $2B$, respectively. In the generalized case $x_e\in \{0,1,\dots,n_e\}$, one can analogously use $s_e\in \{0,1,\dots,Bn_e\}$.

The complete Hashiwokakero cost function is therefore
\begin{equation}
    C_{\mathrm{Hashi}}=C_{\mathrm{deg}}+C_{\mathrm{cross}}+C_{\mathrm{bal}}+C_{\mathrm{cap}}.
\end{equation}
This is still a QUDO formulation, because each new term is the square of a linear expression in bounded integer variables, and therefore it expands into terms of degree at most two.

This construction is exact. If $C_{\mathrm{bal}}+C_{\mathrm{cap}}=0$, let $E_x=\{e\in E:x_e>0\}$ be the active bridge set. Assume by contradiction that the graph $(V,E_x)$ is not connected, and let $S\subseteq V\setminus\{r\}$ be a nonempty subset disconnected from the root. Summing the balance equalities over $S$ gives
\begin{equation}
    \sum_{v\in S}
    \left(
        \sum_{u:\{v,u\}\in E} f_{vu}
        -\sum_{u:\{v,u\}\in E} f_{uv}
    \right)
    =-|S|.
\end{equation}
Therefore, $S$ must receive $|S|$ units of net flow from $V\setminus S$. However, by the capacity constraints, positive flow can cross the cut only through edges with $x_e>0$. Since $S$ is disconnected from $r$ in the active graph, no active edge crosses the cut, which is a contradiction. Therefore, the active bridge graph is connected.

Conversely, if the active bridge graph is connected, choose a spanning tree rooted at $r$ and send one unit of flow from $r$ to every other island along the unique path in that tree. No tree edge carries more than $n-1=B$ units of flow. Since every tree edge is active, $x_e\geq 1$, so the capacity $Bx_e$ is sufficient. Choosing $s_e=Bx_e-f_{ij}-f_{ji}$ makes every balance and capacity equation exact, hence $C_{\mathrm{bal}}+C_{\mathrm{cap}}=0$.

The complete formulation therefore adds $2|E|$ flow variables and $|E|$ capacity slack variables to the local bridge variables. In the standard Hashiwokakero case, the bridge variables are the $|E|$ variables $x_e\in \{0,1,2\}$, the flow variables are the $2|E|$ variables $f_{ij}\in \{0,1,\dots,n-1\}$, and the slack variables are the $|E|$ variables $s_e\in \{0,1,\dots,2(n-1)\}$. If only qudits of fixed dimension $d$ are available, these auxiliary variables can be encoded in base $d$ using the slack encoding described previously. The penalty weights $\lambda_d,\lambda_\times,\lambda_b,\lambda_c$ must be chosen sufficiently large so that violating the structural constraints is never energetically favorable relative to the other terms. Since Hashiwokakero is a feasibility problem, one may simply choose positive weights with the desired relative scaling. We can also generalize the shape of the grid and the connectivity simply by changing the elements of $E$, or equivalently the sets $\delta(v)$.

\section{Tensor Quadratic Unconstrained D-ary Optimization (T-QUDO)}
We have seen the main advantages of the QUDO formulation in terms of reducing the required resources on resolutions. However, we also have seen a lot of limitations in constraint terms. The Tensor Quadratic Unconstrained D-ary Optimization (T-QUDO) is a pairwise categorical generalization of QUDO which can overcome some of these limitations. A convenient form is
\begin{equation}
    C(\vec{x})=\sum_{i=0}^{n-1} U_i(x_i)+\sum_{i<j}^{n-1} V_{ij}(x_i,x_j),
\end{equation}
with $x_i\in\{0,1,\dots,d_i-1\}$. In the tensor notation used in this manuscript, the same information can be written as
\begin{equation}
    C(\vec{x})=\sum_{i\leq j}^{n-1} Q_{i,j,x_i,x_j},
\end{equation}
where $Q_{i,i,a,a}=U_i(a)$ and $Q_{i,j,a,b}=V_{ij}(a,b)$ for $i<j$. We recover Eq.~\eqref{eq: QUDO formulation} when $V_{ij}(a,b)=Q_{ij}ab$ for $i\neq j$ and $U_i(a)=D_{i}a+Q_{ii}a^2$.
\begin{enumerate}
    \item Counting constraint: if we want to determine that there are $N$ variables with non-zero values, i.e. $\sum_{i\in\mathcal{X}} H(x_i) = N$, being $H$ the Heaviside step function, we need a term
    \begin{equation}
        (N-\sum_{i\in\mathcal{X}} H(x_i))^2 = N^2+\sum_{i,j\in\mathcal{X}} H(x_i)H(x_j) -2N\sum_{i\in\mathcal{X}} H(x_i),
    \end{equation}
    which is a T-QUDO expression if $Q_{i,j,x_i,x_j}=2H(x_i)H(x_j)$ if $i\neq j$ (the $2$ is for the symmetry) and $Q_{i,i,x_i,x_i}=H(x_i)^2-2NH(x_i)$ else. If we want the inequality expression, we can easily add slack variables.
    
    \item Non-coincidence constraint: if we want to impose that $x_i= a \Longrightarrow x_j\neq b$ in T-QUDO we use a term
    \begin{equation}
        Q_{i,j,a,b}=\lambda.
    \end{equation}
    
    \item Non-equality constraint: if we want them to satisfy $x_i\neq a \Longrightarrow x_j= b$, we use a term
    \begin{equation}
        Q_{i,j,\overline{a},\overline{b}}=\lambda,
    \end{equation}
    being $\overline{c}$ all possible values for that variable but $c$.
    
    \item Implication constraint: if we have a restriction $x_i=a \Longrightarrow x_j=b$, we use a term
    \begin{equation}
        Q_{i,j,a,\overline{b}}=\lambda.
    \end{equation}
    We can also implement the opposite implication, $x_i\neq a \Longrightarrow x_j\neq b$, with a term
    \begin{equation}
        Q_{i,j,\overline{a}, b}=\lambda.
    \end{equation}
\end{enumerate}

As we can observe, we can easily implement many restrictions that we were unable to implement in the QUDO formulation.

This additional flexibility still has to be encoded if we want to work with binary variables. Let $L_i=\lceil \log_2(d_i)\rceil$ and encode
\begin{equation}
    x_i=\sum_{r=0}^{L_i-1}2^r y_{ir},\qquad y_{ir}\in\{0,1\}.
\end{equation}
For each valid label $a\in\{0,1,\dots,d_i-1\}$, with binary digits $a_r\in\{0,1\}$, define
\begin{equation}
    \chi_{i,a}(y)=\prod_{r=0}^{L_i-1}\left[a_r y_{ir} + (1-a_r)(1-y_{ir})\right].
\end{equation}
Then $\chi_{i,a}(y)=1$ exactly when the binary code of $x_i$ equals $a$, so $\mathbf{1}[x_i=a]=\chi_{i,a}(y)$. Therefore
\begin{equation}
    U_i(x_i)=\sum_{a=0}^{d_i-1} U_i(a)\chi_{i,a}(y),
\end{equation}
and every pairwise categorical term becomes
\begin{equation}
    V_{ij}(x_i,x_j)=\sum_{a=0}^{d_i-1}\sum_{b=0}^{d_j-1} V_{ij}(a,b)\chi_{i,a}(y)\chi_{j,b}(y).
\end{equation}
This is a HOBO term of degree at most $L_i+L_j$. As before, invalid binary codes must be excluded or penalized when $d_i$ is not a power of two. This proves an encoding into HOBO, but not a resource-preserving equivalence. Conversely, a general HOBO instance does not reduce in general to pairwise T-QUDO over the same variables without grouping variables, introducing ancillas, or possibly incurring exponential blow-up.

A quadratic alternative is the one-hot encoding. For each categorical variable, let $z_{i,a}\in\{0,1\}$ and enforce $\sum_{a=0}^{d_i-1} z_{i,a}=1$. Then
\begin{equation}
    C(z)=\sum_{i=0}^{n-1}\sum_{a=0}^{d_i-1} U_i(a)z_{i,a}
    +\sum_{i<j}^{n-1}\sum_{a=0}^{d_i-1}\sum_{b=0}^{d_j-1} V_{ij}(a,b)z_{i,a}z_{j,b}
    +\rho \sum_{i=0}^{n-1}\left(1-\sum_{a=0}^{d_i-1} z_{i,a}\right)^2.
\end{equation}
This is a QUBO formulation which uses more variables, but keeps the degree quadratic.

The last important problem of the formulation is the implementation in the QAOA. In this case, we need to implement the exponential of the cost hamiltonian. To do this, we must use controlled focus phase gates. We define a focus phase gate as
\begin{equation}
    FP(\theta,b)_{lm}= \delta_{lm}\left(\sum_{c\neq b}\delta_{lc} + \delta_{lb}e^{i\theta}\right).
\end{equation}
To impose the $Q_{i,j,a,b}$ term, we need to make a controlled focus phase gate $FP(\theta,b)$, with control in the $i$-th qudit (activated with the state $\ket{a}$) and target in the $j$-th qudit. We can also impose it with an auxiliary qubit in state $\ket{1}$, with a double controlled phase gate, controlled by the $i$-th and $j$-th qudits in states $\ket{a}$ and $\ket{b}$ and target the auxiliary qubit.

We will show the interest of this formulation with the Traveling Salesman Problem, due to its applications, and with the N-Queens problem, because it is not feasible with the QUDO formulation. The T-QUDO of the Knapsack problem is not needed, due to the fact that involves the same variables as in the QUDO case.

\subsection{Traveling Salesman Problem}
The Traveling Salesman Problem (TSP)~\cite{TSP} consists in finding a path between the $V$ nodes of a graph with the lowest possible cost, without repeating any node, and returning to the starting point. We choose as variables $x_t\in\{0,1,\dots,V-1\}$ being the node chosen in the timestep $t$, giving a path $\vec{x}=(x_0,x_1,\dots, x_{V-1})$. This implies the need for $V$ variables, which can be reduced to $V-1$ fixing the first node of the path due to the symmetries. If the path between the $i$-th and $j$-th nodes has a cost $E_{ij}$, the total cost of the path is
\begin{equation}
    C(\vec{x})=\sum_{t=0}^{V-2} E_{x_t,x_{t+1}} + E_{x_{V-1},x_{0}}.
\end{equation}
If two nodes are not connected, its cost will be $E_{ij}=\lambda$.
We can check that this is a particular case of the T-QUDO formulation without the indexes indicating the position of the variables. To impose the restriction of the non-repetition, one option is to use prime numbers. Let $p_0=2,p_1=3,p_2=5,\dots,p_{V-1}$ be $V$ distinct primes and define $r_v=\log p_v$. Then the permutation penalty
\begin{equation}
    C_r^{\mathrm{prime}}=\lambda\left(\sum_{t=0}^{V-1} r_{x_t}-\sum_{v=0}^{V-1} r_v\right)^2
\end{equation}
vanishes only if
\begin{equation}
    \sum_{t=0}^{V-1}\log p_{x_t}=\sum_{v=0}^{V-1}\log p_v.
\end{equation}
Exponentiating both sides gives $\prod_{t=0}^{V-1}p_{x_t}=\prod_{v=0}^{V-1}p_v$, and the uniqueness of prime factorization implies that every vertex appears exactly once. This is a T-QUDO expression because each $r_{x_t}$ is a unary term. However, the prime formulation introduces irrational coefficients and is usually less convenient numerically than the direct pairwise form
\begin{equation}
     C_r = \lambda\sum_{0\leq t<t'\leq V-1}\delta_{x_t,x_{t'}}.
\end{equation}
With this choice, the T-QUDO formulation of the TSP is
\begin{equation}
        C(\vec{x})=\sum_{t=0}^{V-2} E_{x_t,x_{t+1}} + E_{x_{V-1},x_{0}}
        + \lambda\sum_{0\leq t<t'\leq V-1}\delta_{x_t,x_{t'}}.
\end{equation}
We can easily transform this cost function for the TSP generalization with a cost that changes with the timestep. In this case, the cost of the travel from $i$-th vertex to $j$-th vertex at $t$-th timestep is $E_{t,i,j}$, so the T-QUDO cost function is
\begin{equation}
        C(\vec{x})=\sum_{t=0}^{V-2} E_{t,x_t,x_{t+1}} + E_{V-1,x_{V-1},x_{0}}
        + \lambda\sum_{0\leq t<t'\leq V-1}\delta_{x_t,x_{t'}}.
\end{equation}
We can check that this is also a T-QUDO expression.

\subsection{N-Queens}
\begin{figure}[h]
    \centering
    \includegraphics[width=0.5\linewidth]{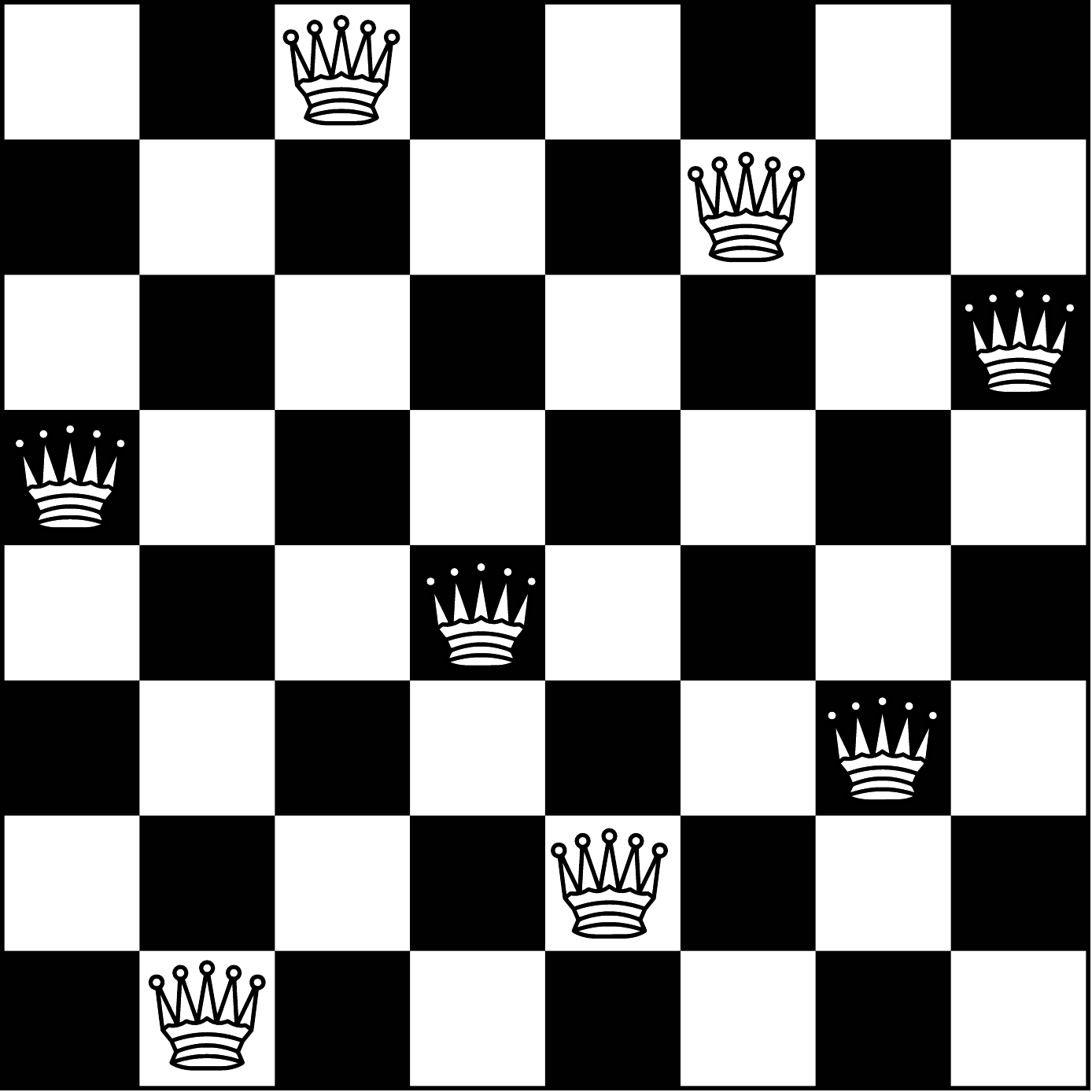}
    \caption{N-Queens solution with a $8\times 8$ chessboard.}
    \label{fig: nqueens}
\end{figure}

The N-Queens problem consists in putting $N$ queens on a chessboard of $N\times N$, in a way that no queen threatens others. A solution is shown in Fig.~\ref{fig: nqueens}. The constraints can be expressed as
\begin{itemize}
    \item Rows condition: there can only be one queen per row.
    \item Columns condition: there can only be one queen per column.
    \item Diagonal condition: there cannot be two queens on the same diagonal.
\end{itemize}
This problem has previously been formulated in a QUBO formulation with more generalizations~\cite{QUBOQueens}. However, if we want to reduce the amount of resources needed, we can use as variables $x_i\in\{0,1,\dots,N-1\}$ the column of the queen in the $i$-th row. We can formulate the T-QUDO problem as
\begin{equation}
    C(\vec{x})=\sum_{0\leq i<j<N}C_{i,j,x_i, x_j},
\end{equation}
being $i$ the row of the first queen and $j$ the row of the second queen. This requires $N$ variables instead of $N^2$.

The cost tensor $C$ must increase the cost of illegal configurations. The non-zero elements are:
\begin{enumerate}
    \item Column condition: if two queens are from different rows, their column must be different. So, if $i\neq j$, then $x_i\neq x_j$:
        \begin{equation}
            C_{i,j,a,a} = \lambda,\qquad \forall 0\leq i<j<N,\ \forall 0\leq a<N.
        \end{equation}
     \item Diagonal condition: if two queens are from different rows, separated by $k$ positions, their column cannot be the same separated by $k$ positions in the same direction. So, if $j=i+k$, then $x_j\neq x_i +k$ nor $x_j\neq x_i -k$:
        \begin{equation}
            C_{i,i+k,a,a+k} = \lambda,\qquad \forall 0\leq i<i+k<N,\ \forall 0\leq a<N-k,
        \end{equation}
        \begin{equation}
            C_{i,i+k,a,a-k} = \lambda,\qquad \forall 0\leq i<i+k<N,\ \forall k\leq a<N.
        \end{equation}
    
\end{enumerate}
All other terms of the tensor are zero. Equivalently, we can write
\begin{equation}
    C(\vec{x})=\lambda\sum_{0\leq i<j<N}\left(\delta_{x_i,x_j} +\delta_{x_i,x_j+(j-i)}+\delta_{x_i,x_j-(j-i)}\right),
\end{equation}
with the convention that $\delta_{a,b}=0$ when $b\notin\{0,1,\dots,N-1\}$. This is equivalent to the indicator form $\lambda\sum_{i<j}\left[\mathbf{1}[x_i=x_j]+\mathbf{1}[x_i-x_j=j-i]+\mathbf{1}[x_j-x_i=j-i]\right]$.

\subsection{Kakuro}
\begin{figure}[h]
    \centering
    \includegraphics[width=\linewidth]{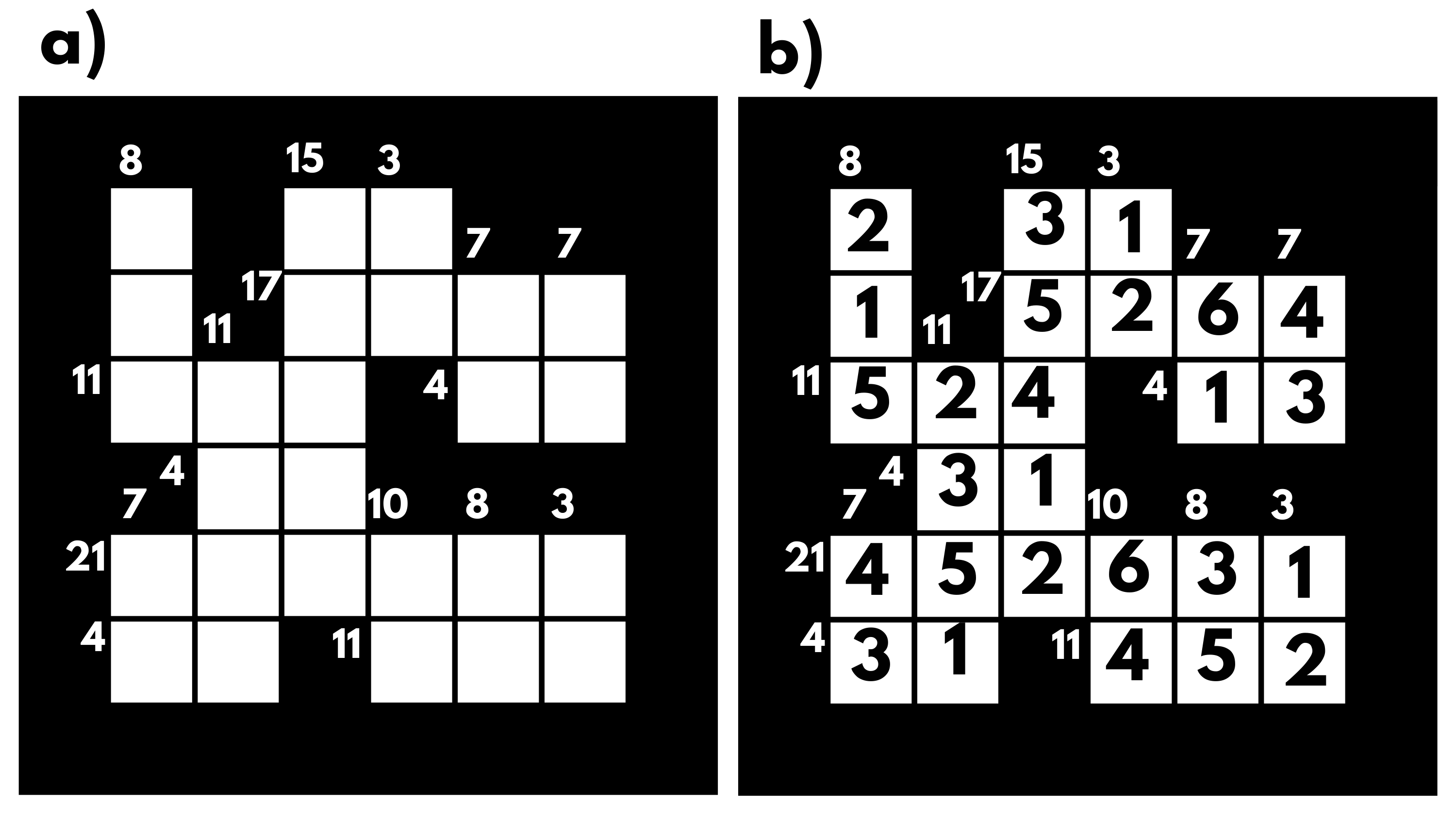}
    \caption{a) Kakuro problem for a $6\times 6$ board, b) solution for a $6\times 6$ Kakuro.}
    \label{fig: kakuro}
\end{figure}

This game consists of a board with $N\times N$ cells, some of them not available (black cells) and others available (white cells) to have a number from $1$ to $M$. Each portion of a row or column has a number that tells the sum of the numbers of that portion of a row/column. Additionally, each number can appear only once in each portion of a row and column. An example of a solution is shown in Fig.~\ref{fig: kakuro}.

It is convenient to distinguish the qudit labels from the physical cell values. For each white cell, let $x_{ij}\in\{0,1,\dots,M-1\}$ be the qudit label and $y_{ij}=x_{ij}+1\in\{1,2,\dots,M\}$ the number written in the cell. The $k$-th portion of the $i$-th row is defined by the cells in $\mathcal{R}_{ik}$ and the $k$-th portion of the $j$-th column is defined by the cells in $\mathcal{C}_{jk}$. The portion $\mathcal{R}_{ik}$ has an associated sum number $S^r_{ik}$ and the portion $\mathcal{C}_{jk}$ has an associated sum number $S^c_{jk}$. If we have qudits of dimension $M$, we need only as variables the white cells, at most $N^2$. Otherwise, we can use more variables with the encoding previously shown.

With these definitions, we can define the T-QUDO formulation. First, we define the term that implements the sum constraint. We know that $\sum_{i',j'\in \mathcal{R}_{ik}} y_{i'j'}=S^r_{ik}\ \forall i,k$, and $\sum_{i',j'\in \mathcal{C}_{jk}} y_{i'j'}=S^c_{jk}\ \forall j,k$, so the constraint term is
\begin{equation}
    C_s = \lambda_s\sum_{i=0}^{N-1}\sum_{k}\left(\sum_{i',j'\in \mathcal{R}_{ik}} y_{i'j'} - S^r_{ik}\right)^2+\lambda_s\sum_{j=0}^{N-1}\sum_{k}\left(\sum_{i',j'\in \mathcal{C}_{jk}} y_{i'j'} - S^c_{jk}\right)^2.
\end{equation}
This is a QUDO term in the numeric variables $y_{ij}$. The second restriction, the non-repetition one, is pairwise and therefore naturally T-QUDO for $M>2$. It can be written as
\begin{equation}
    \begin{gathered}
        C_r=\lambda_r\sum_{i=0}^{N-1}\sum_{k}\sum_{\substack{(i',j'),(l',m')\in \mathcal{R}_{ik}\\ (i',j')<(l',m')}}\delta_{x_{i'j'},x_{l'm'}}+\\
        +\lambda_r\sum_{j=0}^{N-1}\sum_{k}\sum_{\substack{(i',j'),(l',m')\in \mathcal{C}_{jk}\\ (i',j')<(l',m')}}\delta_{x_{i'j'},x_{l'm'}}.
    \end{gathered}
\end{equation}
where we can equivalently use $\delta_{y_{i'j'},y_{l'm'}}$ because $y_{ij}=x_{ij}+1$. The total cost function is
\begin{equation}
    C(x)=C_s+C_r.
\end{equation}
We can check that the sum constraint is QUDO while the non-repetition term is T-QUDO.

\subsection{Inshi no Heya}
\begin{figure}[h]
    \centering
    \includegraphics[width=\linewidth]{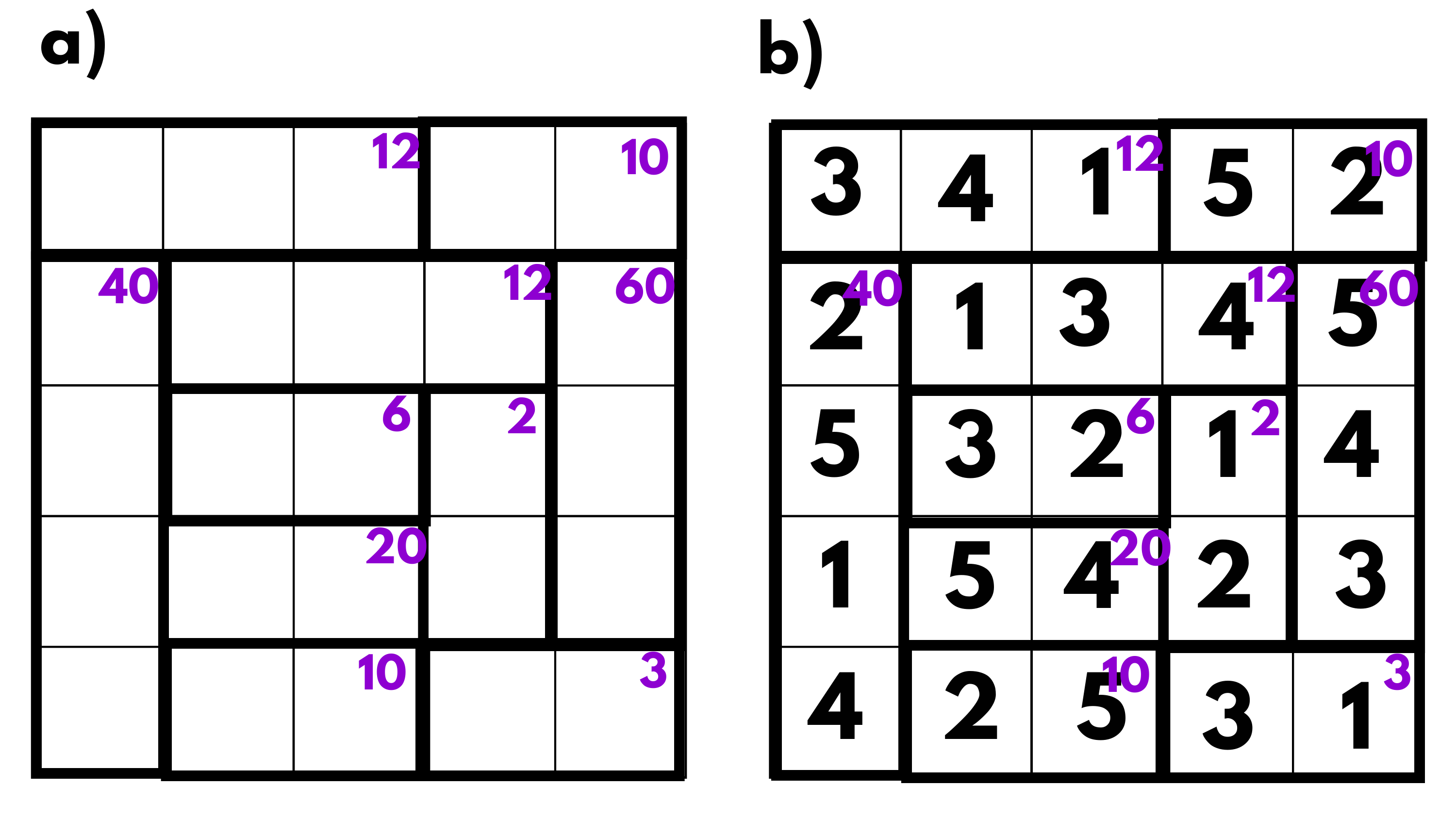}
    \caption{a) Inshi no Heya problem for a $5\times 5$ board, b) Solution for the $5\times 5$ Inshi no Heya.}
    \label{fig: Inshi no Heya}
\end{figure}

The Inshi no Heya puzzle is similar to a Latin-square puzzle, but in this case the $N\times N$ board is divided into regions (originally rectangular, but we can generalize it to arbitrary shapes). We need to give a number from $1$ to $N$ to each cell in the board, without repeating numbers in the same row or column. The standard room rule is multiplicative: if $\mathcal{R}_k$ is the $k$-th region with clue $S_k$, the numbers placed in that region must satisfy $\prod_{c\in \mathcal{R}_k} y_c=S_k$. An example is shown in Fig.~\ref{fig: Inshi no Heya}.

We use physical values $y_{ij}\in\{1,2,\dots,N\}$ as the value in the cell of the $i$-th row and $j$-th column. Let $P_N$ be the set of primes not larger than $N$. For each $q\in P_N$, let $\nu_q(a)$ be the exponent of $q$ in the prime factorization of $a$, and factor each region clue as
\begin{equation}
    S_k=\prod_{q\in P_N} q^{e_{k,q}}.
\end{equation}
Then the multiplicative region rule is equivalent to
\begin{equation}
    \prod_{c\in \mathcal{R}_k} y_c=S_k
    \qquad\Longleftrightarrow\qquad
    \sum_{c\in \mathcal{R}_k}\nu_q(y_c)=e_{k,q}\quad \forall q\in P_N.
\end{equation}
Therefore, the product constraint can be imposed with
\begin{equation}
    C_p=\lambda_p\sum_{k}\sum_{q\in P_N}\left(\sum_{c\in \mathcal{R}_k}\nu_q(y_c)-e_{k,q}\right)^2.
\end{equation}
Each $\nu_q(y_c)$ is a unary function of a categorical variable, so squaring only creates unary and pairwise terms. The Latin-square restrictions can be written as
\begin{equation}
    C_l=\lambda_l\sum_{i=0}^{N-1}\sum_{0\leq j<j'<N}\delta_{y_{ij},y_{ij'}}
    +\lambda_l\sum_{j=0}^{N-1}\sum_{0\leq i<i'<N}\delta_{y_{ij},y_{i'j}}.
\end{equation}
Because each row and column contains $N$ cells and the values are in $\{1,2,\dots,N\}$, non-repetition implies that every value appears exactly once in each row and in each column. The total cost function is then
\begin{equation}
    C(y)=C_p+C_l.
\end{equation}
This is a T-QUDO formulation of the standard multiplicative version of Inshi no Heya.

\section{Higher-Order Binary Optimization (HOBO)}
We have seen the formalisms to work with qudits, but they are not directly implementable in qubit devices. Also, these formalisms allow us to work only with interactions between pairs of variables, but not more complex ones. These are good reasons to explore higher-order formalisms. The Higher-Order Binary Optimization (HOBO) is the simplest one that we can formulate, with a multilinear pseudo-Boolean cost function of order at most $m$
\begin{equation}
    C(\vec{y})=\sum_{S\subseteq \{0,1,\dots,n-1\},|S|\leq m} q_S \prod_{i\in S} y_i,
\end{equation}
where $y_i\in\{0,1\}$. Because $y_i^2=y_i$, every polynomial in binary variables can be reduced to this multilinear form. Due to the fact that QUBO is the special case $|S|\leq 2$, HOBO can express encoded QUBO and T-QUDO models. So, the main advantage of this formulation is the additional possible interactions. It has been previously studied and implemented in~\cite{HOBO_rail,Opt_HOBO,QUBO_HOBO_TSP}. We will see its potential with the popular game Peg Solitaire.

The interesting thing is that pairwise categorical T-QUDO interactions can be encoded into HOBO interactions by binarizing the qudit labels, as shown in the previous section. In the second QAOA implementation of the proposed T-QUDO, we use two qudit controls to apply a phase gate to an ancilla qubit. However, we can change each of the qudit controls to a set of qubit controls, which creates the binary representation of the states of the qudit. In this way, if we had a qudit of dimension $d_0$ and another qudit of dimension $d_1$ which controls the phase gate, with control states $\ket{a}$ and $\ket{b}$, we substitute them with $\lceil\log_2(d_0)\rceil$ and $\lceil\log_2(d_1)\rceil$ qubits, respectively, with control states given by the binary representation of $\ket{a}$ and $\ket{b}$. This yields a HOBO realization of the encoded T-QUDO interaction, but not a resource-preserving equivalence in general.

\subsection{Peg Solitaire}
\begin{figure}[h]
    \centering
    \includegraphics[width=\linewidth]{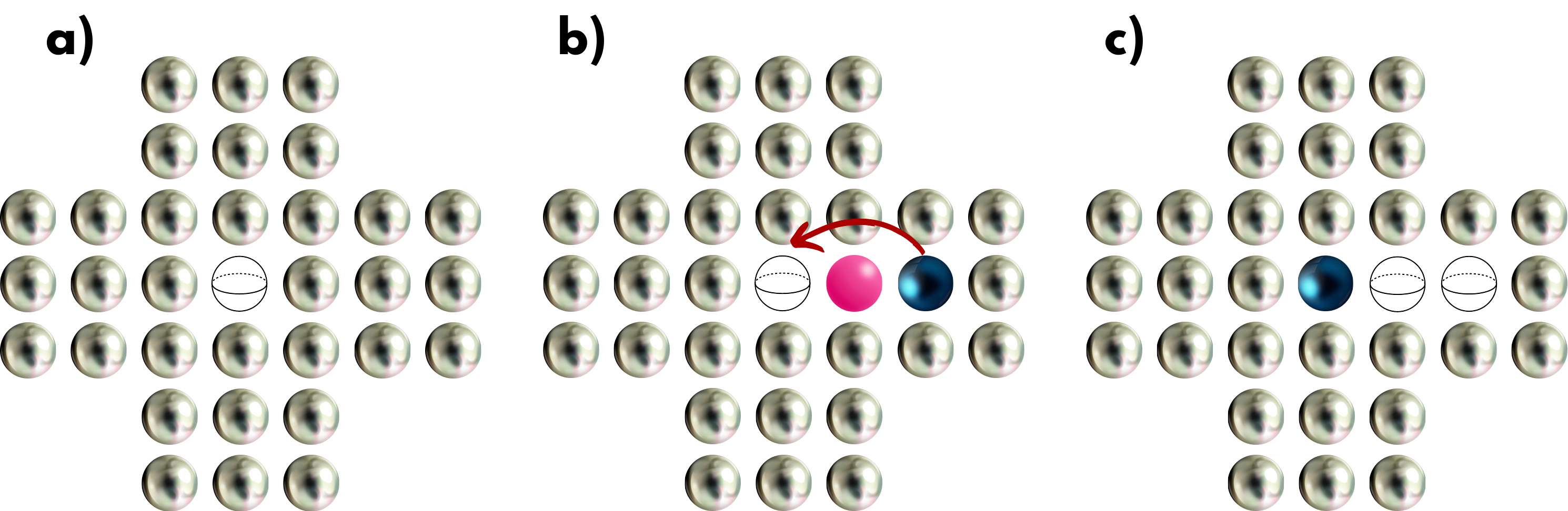}
    \caption{a) Initial configuration of the Peg Solitaire, b) movement of the green ball across the red ball, c) result from the movement.}
    \label{fig: Peg solitaire}
\end{figure}

Peg Solitaire consists of a board with a certain shape, with a set of balls in each cell but one, as shown in Fig.~\ref{fig: Peg solitaire} a. In each turn, the player can make a movement. Each movement consists in moving a ball from its cell to an empty cell by jumping over an adjacent occupied cell in one of the four cardinal directions. If the ball starts in the cell $(i,j)$, there is a ball in the intermediate cell and it lands two positions away in the same direction. An example is shown in Fig.~\ref{fig: Peg solitaire} b. After the movement, the jumped-over ball disappears from the board, as shown in Fig.~\ref{fig: Peg solitaire} c. The objective is to make a set of moves that end in having a unique ball in the starting empty cell and no other balls.

We present a HOBO formulation on a board with valid cells in a set $\mathcal{C}$, with $M=|\mathcal{C}|$. For a square board, $M=N^2$. The binary variable $x_{c,t}$ indicates the state of cell $c\in\mathcal{C}$ at configuration $t$: it is $1$ if the cell is occupied and $0$ otherwise. If the initial configuration is indexed by $t=0$, then we need $T=M-1$ configurations, indexed by $t=0,\dots,T-1$, to remove $M-2$ pegs. If we fix the first and last configurations, the number of free state variables is $M(T-2)=M(M-3)$.

If $c_\star$ is the initially empty cell, we impose
\begin{equation}
    x_{c,0}=1-\delta_{c,c_\star},\qquad x_{c,T-1}=\delta_{c,c_\star}.
\end{equation}

To describe legal moves, we define the valid action set
\begin{equation}
    \begin{aligned}
        A=\{(s,m,d):&\ m=s+\Delta,\ d=s+2\Delta,\\
        &\ \Delta\in\{(1,0),(-1,0),(0,1),(0,-1)\},\ s,m,d\in\mathcal{C}\}.
    \end{aligned}
\end{equation}
For each move time $t=0,\dots,T-2$ and each action $\alpha\in A$, we introduce a binary variable $a_{\alpha,t}$. This adds $|A|(M-2)$ action variables. If we ignore boundary filtering and simply allow four directions per cell, we get the upper bound $4M(M-2)$ action variables, and therefore the total number of free variables is bounded by
\begin{equation}
    M(M-3)+4M(M-2)=5M^2-11M.
\end{equation}
For a square board, this becomes $5N^4-11N^2$.

The peg-count condition is
\begin{equation}
    C_q=\lambda_q\sum_{t=0}^{T-1}\left((M-1-t)-\sum_{c\in\mathcal{C}}x_{c,t}\right)^2.
\end{equation}
We can also enforce that exactly one action is taken at each move time with
\begin{equation}
    C_a=\lambda_a\sum_{t=0}^{T-2}\left(1-\sum_{\alpha\in A}a_{\alpha,t}\right)^2.
\end{equation}

If $\alpha=(s,m,d)$, then a valid move at time $t$ must satisfy the precondition
\begin{equation}
    x_{s,t}=1,\qquad x_{m,t}=1,\qquad x_{d,t}=0,
\end{equation}
and the postcondition
\begin{equation}
    x_{s,t+1}=0,\qquad x_{m,t+1}=0,\qquad x_{d,t+1}=1.
\end{equation}
Define
\begin{equation}
    P^{\mathrm{pre}}_{\alpha,t}=x_{s,t}x_{m,t}(1-x_{d,t}),
\end{equation}
\begin{equation}
    P^{\mathrm{post}}_{\alpha,t}=(1-x_{s,t+1})(1-x_{m,t+1})x_{d,t+1}.
\end{equation}
Then the movement penalty can be written as
\begin{equation}
    C_m=\lambda_m\sum_{t=0}^{T-2}\sum_{\alpha\in A} a_{\alpha,t}\left[\left(P^{\mathrm{pre}}_{\alpha,t}-1\right)^2+\left(P^{\mathrm{post}}_{\alpha,t}-1\right)^2\right].
\end{equation}
For binary variables, $\left(P^{\mathrm{pre}}_{\alpha,t}-1\right)^2=1-P^{\mathrm{pre}}_{\alpha,t}$ and $\left(P^{\mathrm{post}}_{\alpha,t}-1\right)^2=1-P^{\mathrm{post}}_{\alpha,t}$, so after multiplying by $a_{\alpha,t}$ and using idempotence, the highest degree is $4$, not $7$.

If we want to keep an explicit continuity penalty, we can use
\begin{equation}
    C_c=\lambda_c\sum_{t=1}^{T-1}\left((M-3)-\sum_{c\in\mathcal{C}}\left[x_{c,t}x_{c,t-1}+(1-x_{c,t})(1-x_{c,t-1})\right]\right)^2.
\end{equation}
This term is a helper penalty: once $C_a$ and $C_m$ enforce exactly one valid move per time, $C_c$ is redundant, although it may still help numerically in some implementations.

The global HOBO formulation can therefore be written as
\begin{equation}
    C(x,a)=C_q+C_a+C_m+C_c,
\end{equation}
with the understanding that $C_c$ is optional. The correct combination of movements which solves the problem has cost zero.

\section{Conclusions}
We have presented and studied the QUDO, T-QUDO and HOBO formulations, with their advantages and limitations. We also applied them for the knapsack problem, the traveling salesman problem, and five interesting combinatorial games. As we show, many problems can be formulated in an easier and more efficient way with these more complex formalisms. However, it may be difficult if we do not choose the correct formulation. Moreover, QUDO can be encoded into QUBO and T-QUDO can be encoded into HOBO or QUBO through standard binary or one-hot encodings, although these translations may require penalties for invalid states and do not preserve resources in general. Moreover, we show that these formalisms are implementable in the QAOA with small modifications.

Future lines of research may include a deeper analysis of the details of implementation, optimization of interaction types, the design of efficient hardware architectures, and the application to other problems.

\section*{Acknowledgment}
This work has been developed in the `When Physics Becomes Science' project of \href{https://www.youtube.com/@whenphysics}{When Physics}, an initiative to recover the original vision of science.

\bibliographystyle{unsrt}  
\bibliography{references}

@PREAMBLE{
 "\providecommand{\noopsort}[1]{}" 
 # "\providecommand{\singleletter}[1]{#1}%" 
}

@misc{Comparing_Nqueens,
      title={Comparing Python, Go, and C++ on the N-Queens Problem}, 
      author={Pascal Fua and Krzysztof Lis},
      year={2020},
      eprint={2001.02491},
      archivePrefix={arXiv},
      primaryClass={cs.MS},
      url={https://arxiv.org/abs/2001.02491}, 
}

@misc{Quantum_Queens,
      title={A Quantum Approach to solve N-Queens Problem}, 
      author={Santhosh G S and Piyush Joshi and Ayan Barui and Prasanta K. Panigrahi},
      year={2023},
      eprint={2312.16312},
      archivePrefix={arXiv},
      primaryClass={quant-ph},
      url={https://arxiv.org/abs/2312.16312}, 
}

@misc{QAOA,
      title={A Quantum Approximate Optimization Algorithm}, 
      author={Edward Farhi and Jeffrey Goldstone and Sam Gutmann},
      year={2014},
      eprint={1411.4028},
      archivePrefix={arXiv},
      primaryClass={quant-ph},
      url={https://arxiv.org/abs/1411.4028}, 
}

@misc{QUBO,
      title={A Tutorial on Formulating and Using QUBO Models}, 
      author={Fred Glover and Gary Kochenberger and Yu Du},
      year={2019},
      eprint={1811.11538},
      archivePrefix={arXiv},
      primaryClass={cs.DS},
      url={https://arxiv.org/abs/1811.11538}, 
}

@misc{QUBOQueens,
      title={A QUBO Formulation for the Generalized LinkedIn Queens and Takuzu/Tango Game}, 
      author={Alejandro Mata Ali and Edgar Mencia},
      year={2024},
      eprint={2410.06429},
      archivePrefix={arXiv},
      primaryClass={quant-ph},
      url={https://arxiv.org/abs/2410.06429}, 
}

@misc{QUBO_HOBO_TSP,
      title={Beyond QUBO and HOBO formulations, solving the Travelling Salesman Problem on a quantum boson sampler}, 
      author={Daniel Goldsmith and Joe Day-Evans},
      year={2024},
      eprint={2406.14252},
      archivePrefix={arXiv},
      primaryClass={quant-ph},
      url={https://arxiv.org/abs/2406.14252},
}

@article{Symbolic_Queens,
title = {An incremental approach to the n-queen problem with polynomial time},
journal = {Journal of King Saud University - Computer and Information Sciences},
volume = {35},
number = {3},
pages = {1-7},
year = {2023},
issn = {1319-1578},
doi = {https://doi.org/10.1016/j.jksuci.2023.02.002},
url = {https://www.sciencedirect.com/science/article/pii/S1319157823000319},
author = {Bouneb Zine {El Abidine}}
}

@article{Survey_Queens,
title = {A survey of known results and research areas for n-queens},
journal = {Discrete Mathematics},
volume = {309},
number = {1},
pages = {1-31},
year = {2009},
issn = {0012-365X},
doi = {https://doi.org/10.1016/j.disc.2007.12.043},
url = {https://www.sciencedirect.com/science/article/pii/S0012365X07010394},
author = {Jordan Bell and Brett Stevens}
}

@Article{Non_conflict_Queens,
AUTHOR = {Moghimi, Omid and Amini, Amin},
TITLE = {A Novel Approach for Solving the N-Queen Problem Using a Non-Sequential Conflict Resolution Algorithm},
JOURNAL = {Electronics},
VOLUME = {13},
YEAR = {2024},
NUMBER = {20},
ARTICLE-NUMBER = {4065},
URL = {https://www.mdpi.com/2079-9292/13/20/4065},
ISSN = {2079-9292},
DOI = {10.3390/electronics13204065}
}

@InProceedings{GPU_Queens,
author="Zhang, Tao
and Shu, Wei
and Wu, Min-You",
editor="Temam, Olivier
and Yew, Pen-Chung
and Zang, Binyu",
title="Optimization of N-Queens Solvers on Graphics Processors",
booktitle="Advanced Parallel Processing Technologies",
year="2011",
publisher="Springer Berlin Heidelberg",
address="Berlin, Heidelberg",
pages="142--156",
isbn="978-3-642-24151-2"
}

@article{Hashi_NP,
title = {Hashiwokakero is NP-complete},
journal = {Information Processing Letters},
volume = {109},
number = {19},
pages = {1145-1146},
year = {2009},
issn = {0020-0190},
doi = {https://doi.org/10.1016/j.ipl.2009.07.017},
url = {https://www.sciencedirect.com/science/article/pii/S002001900900235X},
author = {Daniel Andersson}
}

@misc{Bench_Hashi,
      title={Benchmark Instances and Branch-and-Cut Algorithm for the Hashiwokakero Puzzle}, 
      author={Leandro C. Coelho and Gilbert Laporte and Arinei Lindbeck and Thibaut Vidal},
      year={2019},
      eprint={1905.00973},
      archivePrefix={arXiv},
      primaryClass={cs.DM},
      url={https://arxiv.org/abs/1905.00973}, 
}

@article{Model_Peg,
title = {Modelling and solving English Peg Solitaire},
journal = {Computers \& Operations Research},
volume = {33},
number = {10},
pages = {2935-2959},
year = {2006},
note = {Part Special Issue: Constraint Programming},
issn = {0305-0548},
doi = {https://doi.org/10.1016/j.cor.2005.01.018},
url = {https://www.sciencedirect.com/science/article/pii/S0305054805000195},
author = {Christopher Jefferson and Angela Miguel and Ian Miguel and S. Armagan Tarim}
}

@Article{HOBO_rail,
author={Domino, Krzysztof
and Kundu, Akash
and Salehi, {\"O}zlem
and Krawiec, Krzysztof},
title={Quadratic and higher-order unconstrained binary optimization of railway rescheduling for quantum computing},
journal={Quantum Information Processing},
year={2022},
month={Sep},
day={29},
volume={21},
number={9},
pages={337},
issn={1573-1332},
doi={10.1007/s11128-022-03670-y},
url={https://doi.org/10.1007/s11128-022-03670-y}
}

@INPROCEEDINGS {Opt_HOBO,
author = { Verchere, Zoe and Elloumi, Sourour and Simonetto, Andrea },
booktitle = { 2023 IEEE International Conference on Quantum Computing and Engineering (QCE) },
title = {{ Optimizing Variational Circuits for Higher-Order Binary Optimization }},
year = {2023},
volume = {},
ISSN = {},
pages = {19-25},
doi = {10.1109/QCE57702.2023.00011},
url = {https://doi.ieeecomputersociety.org/10.1109/QCE57702.2023.00011},
publisher = {IEEE Computer Society},
address = {Los Alamitos, CA, USA},
month =sep}

@article{Qudits,
   title={Qudits and High-Dimensional Quantum Computing},
   volume={8},
   ISSN={2296-424X},
   url={http://dx.doi.org/10.3389/fphy.2020.589504},
   DOI={10.3389/fphy.2020.589504},
   journal={Frontiers in Physics},
   publisher={Frontiers Media SA},
   author={Wang, Yuchen and Hu, Zixuan and Sanders, Barry C. and Kais, Sabre},
   year={2020},
   month=nov }

@article{QAOA_qudit,
   title={Quantum approximate optimization algorithm for qudit systems},
   volume={107},
   ISSN={2469-9934},
   url={http://dx.doi.org/10.1103/PhysRevA.107.062410},
   DOI={10.1103/physreva.107.062410},
   number={6},
   journal={Physical Review A},
   publisher={American Physical Society (APS)},
   author={Deller, Yannick and Schmitt, Sebastian and Lewenstein, Maciej and Lenk, Steve and Federer, Marika and Jendrzejewski, Fred and Hauke, Philipp and Kasper, Valentin},
   year={2023},
   month=jun }

@book{knapsack,
  title={Knapsack problems: algorithms and computer implementations},
  author={Martello, Silvano and Toth, Paolo},
  year={1990},
  publisher={John Wiley \& Sons, Inc.}
}

@BOOK{TSP,
  author={Applegate, David L. and Bixby, Robert E. and Chvátal, Vašek and Cook, William J.},
  booktitle={The Traveling Salesman Problem: A Computational Study},
  year={2007},
  volume={},
  number={},
  pages={},
  doi={}}

@article{Unbalanced,
   title={Unbalanced penalization: a new approach to encode inequality constraints of combinatorial problems for quantum optimization algorithms},
   volume={9},
   ISSN={2058-9565},
   url={http://dx.doi.org/10.1088/2058-9565/ad35e4},
   DOI={10.1088/2058-9565/ad35e4},
   number={2},
   journal={Quantum Science and Technology},
   publisher={IOP Publishing},
   author={Montañez-Barrera, J A and Willsch, Dennis and Maldonado-Romo, A and Michielsen, Kristel},
   year={2024},
   month=apr, pages={025022} }

\end{document}